\documentclass[12pt]{article}
\usepackage{epsfig}
\newcommand\bea{\begin{eqnarray}}
\newcommand\eea{\end{eqnarray}}
\begin{document}
\thispagestyle{empty}
\bibliographystyle{unsrt}
\setlength{\baselineskip}{18pt}
\parindent 24pt
\vspace{50pt}

\begin{center}{
{\Large {{\bf Dissipative tunneling through a parabolic potential\\
in the Lindblad theory of open quantum systems }} } \vskip 1truecm
A. Isar${\dagger\ddagger}^{(a)}$, A. Sandulescu${\dagger}$ and W.
Scheid${\ddagger}$\\ $\dagger${\it Department of Theoretical
Physics, Institute of Physics and Nuclear Engineering\\
Bucharest-Magurele,
Romania }\\
 $\ddagger${\it Institut f\"ur Theoretische Physik der
Justus-Liebig-Universit\"at \\ Giessen, Germany }\\ }
\end{center}

\begin{abstract}
By using the Lindblad theory for open quantum systems, an
analytical expression of the tunneling probability
through an inverted parabola is obtained. This
penetration probability depends on the environment
coefficients. It is shown that the tunneling probability
increases with the dissipation and the temperature of the
thermal bath.

\end{abstract}


(a) e-mail address: isar@theory.nipne.ro

\section{Introduction}

Quantum tunneling with dissipation has been intensively investigated
in the last two decades [1--12]. Very interesting is the discussion
whether the dissipation suppresses or enhances the quantum
tunneling. Caldeira and Leggett \cite{cald1,cald2} concluded that
dissipation tends to suppress quantum tunneling. Using a different
method, Schmid obtained similar results in Ref. \cite{schm}. Widom
and Clark \cite{wid} considered a parabolic potential barrier and
found that dissipation enhances tunneling. Bruinsma and Bak
\cite{bru} also considered tunneling through a barrier and found
that at zero temperature the tunneling rate can be either increased
or decreased by dissipation. Leggett \cite{leg2} considered
tunneling in the presence of an arbitrary dissipation mechanism and
found that, normally, dissipation impedes tunneling, but he also
found an anomalous case in which dissipation assists the tunneling
process. Razavy \cite{raz1} considered tunneling in a symmetric
double-well potential and concluded that dissipation can inhibit or
suppress tunneling. Fujikawa {\it et al.} \cite{fuji} also
considered tunneling in a double-well potential and found an
enhancement of tunneling. Harris \cite{har} calculated the tunneling
out of a metastable state in the presence of an environment at zero
temperature and found that quantum tunneling is enhanced by
dissipation. In \cite{yu}, Yu considered the tunneling problem in an
Ohmic dissipative system with inverted harmonic potential and he
showed that while the dissipation  tends to suppress the tunneling,
the Brownian motion tends to enhance it. In a series of papers
\cite{anke1,anke2}, Ankerhold, Grabert and Ingold have studied real
time dynamics of a quantum system with a potential barrier coupled
to a heat bath environment, employing the path integral approach.
The conclusion drawn from these papers is that different authors
have studied different problems using different methods. They
obtained results which in many cases present an evident
contradiction.

In the present paper we study the tunneling in the presence of a
dissipative environment in the framework of the Lindblad theory for
open quantum systems, based on completely positive dynamical
semigroups [23--25]. We extend the work done in some previous papers
\cite{stef,anto}. In \cite{anto} a similar problem was treated by
using the path integral method and numerical calculations. Our study
can be applied to problems of nuclear fragmentation, fission and
fusion, considered as a tunneling process through the nuclear
barrier defined in the space of collective coordinates, like charge
and mass asymmetry or the distance between the fission fragments.

For the inverted harmonic potential, the tunneling problem in the
framework of the Lindblad theory can be solved exactly. In Sec. 2 we
write the basic equations of the Lindblad theory for open quantum
systems and give results for the coordinate and momentum expectation
values and variances for the damped inverted harmonic oscillator.
Then in Sec. 3 we consider the penetration of a Gaussian wave packet
through the potential barrier and define the penetration
probability. In Sec. 4 we analyze its dependence on various
dimensionless parameters which enter the theory and show that the
probability increases with the dissipation and the temperature of
the thermal bath. A summary is given in Sec. 5.

\section{Quantum Markovian master equation in \\ Lindblad theory}

The simplest dynamics for an open system which describes an
irreversible process is a semigroup of transformations introducing a
preferred direction in time [23--25]. In Lindblad's axiomatic
formalism of introducing dissipation in quantum mechanics, the usual
von Neumann-Liouville equation ruling the time evolution of closed
quantum systems is replaced by the following quantum master equation
for the density operator $ \rho(t)$ in the Schr\"odinger picture
\cite{l1}, which is the most general Markovian evolution equation
preserving the positivity, hermiticity and unit trace of $\rho$:
\bea{d\rho(t)\over dt}=-{i\over\hbar}[H,\rho(t)]+{1\over 2\hbar}
\sum_{j}(2V_{j}\rho(t)V_{j}^\dagger- V_{j}^\dagger
V_j\rho(t)-\rho(t)V_j^\dagger V_{j}).\label{lineq}\eea Here $ H$ is
the Hamiltonian operator of the system; $     V_{j},$ $
V_{j}^\dagger $ are bounded operators on the Hilbert space of the
Hamiltonian and model the effect of the environment. We make the
basic assumption that the general form (\ref{lineq}) of the master
equation with a bounded generator is also valid for an unbounded
generator.

As usual, we define the two possible environment
operators $    V_{1}$ and $    V_{2},$ which are assumed
linear in momentum $ p$ and coordinate $ q,$ as follows
\cite{rev,ss}: \bea V_{j}=a_{j} p+b_{j} q,~j=1,2,
\label{oper}\eea with $a_{j},b_{j}$ complex numbers. The
Hamiltonian $     H$ is chosen of the general form \bea
H=     H_{0}+{\mu \over 2}( q p+     p     q), ~~~
H_{0}={1\over 2m}
     p^2+U(     q),  \label{ham}   \eea
where $U(q)$ is the potential.
With these choices and with the notations
\bea D_{qq}={\hbar\over 2}\sum_{j=1,2}{\vert a_{j}\vert}^2,
  D_{pp}={\hbar\over 2}\sum_{j=1,2}{\vert b_{j}\vert}^2,
D_{pq}=D_{qp}=-{\hbar\over 2}{\rm Re}\sum_{j=1,2}a_{j}^*b_{j},
\lambda=-{\rm Im}\sum_{j=1,2}a_{j}^*b_{j},\label{coef}\eea where
$a_j^*$ and $b_j^*$ denote the complex conjugate of $a_j$ and $b_j,$
respectively, the master equation (\ref{lineq}) takes the following
form \cite{rev,ss}: \bea   {d    \rho \over dt}=-{i\over \hbar}[
H_{0},    \rho]- {i\over 2\hbar}(\lambda +\mu) [     q,    \rho p+ p
\rho]+{i\over 2\hbar}(\lambda -\mu)[p,
    \rho      q+     q    \rho]  \nonumber\\
  -{D_{pp}\over {\hbar}^2}[     q,[q,    \rho]]-{D_{qq}\over {\hbar}^2}
[     p,[     p,    \rho]]+{D_{pq}\over {\hbar}^2}([ q,[
p, \rho]]+ [     p,[     q,    \rho]]). ~~~~\label{mast}
\eea Here the quantum diffusion coefficients
$D_{pp},D_{qq},$ $D_{pq}$ and the dissipation constant
$\lambda$ satisfy the following fundamental constraints
\cite{rev,ss}: $  D_{pp}>0, D_{qq}>0$ and \bea
D_{pp}D_{qq}-D_{pq}^2\ge {{\lambda}^2{\hbar}^2\over 4}.
\label{ineq}  \eea In the particular case when the
asymptotic state is a Gibbs state \bea   \rho
_G(\infty)=e^{-{H_0\over kT}}/ {\rm Tr}e^{-{H_0\over
kT}},  \label{gib}  \eea the coefficients for a harmonic
oscillator potential have the following form
\cite{rev,ss}: \bea D_{pp}={\lambda+\mu\over 2}\hbar
m\omega\coth{\hbar\omega\over 2kT},
~~D_{qq}={\lambda-\mu\over 2}{\hbar\over
m\omega}\coth{\hbar\omega\over 2kT}, ~~D_{pq}=0,
\label{coegib} \eea where $T$ is the temperature of the
thermal bath. The fundamental constraint (\ref{ineq}) is
satisfied only if $\lambda>\mu.$

In the following we denote by $\sigma_{AA}$ the dispersion
(variance) of the operator $     A$, i.e. $\sigma_{AA}=<
A^2>-< A>^2,$ where $<     A>\equiv\sigma_A={\rm Tr}(    \rho
A)$ is the expectation value of the operator $A$ and ${\rm Tr}
\rho=1.$ By $\sigma_{AB}=1/2< A B+ B A>-< A>< B>$ we denote the
correlation of the operators $ A$ and $ B.$

From the master equation (\ref{mast}) we obtain the following
equations of motion for the expectation values and variances of
coordinate and momentum: \bea{d\sigma_{q}(t)\over
dt}=-(\lambda-\mu)\sigma_{q}(t)+{1\over m}\sigma_{p}
(t),\label{eqmo1}\eea \bea{d\sigma_{p}(t)\over dt}=-<{dU( q)\over
d     q}>- (\lambda+\mu)\sigma_{p}(t) \label{eqmo2}\eea and,
respectively, \bea{d\sigma_{qq}(t)\over
dt}=-2(\lambda-\mu)\sigma_{qq}(t)+{2\over m}
\sigma_{pq}(t)+2D_{qq},\label{eqmo3}\eea \bea{d\sigma_{pp}\over
dt}=-2(\lambda+\mu)\sigma_{pp}(t)-
<{dU(     q)\over d     q}     p+     p
{dU(     q)\over d     q}>
+2<{dU(     q)\over d     q}>\sigma_p(t)
+2D_{pp}, \label{eqmo4}\eea
\bea{d\sigma_{pq}(t)\over dt}=-
<{dU(     q)\over d     q}     q>+
<{dU(     q)\over d     q}>\sigma_q(t)
+{1\over m}\sigma_{pp}(t)
-2\lambda\sigma_{pq}(t)+2D_{pq}.\label{eqmo5}\eea

For the harmonic oscillator with the potential $U(q)=m\omega^2
q^2/2,$ the solutions of these equations of motion are obtained in
Refs. \cite{rev,ss}. In this paper we consider the tunneling through
a potential barrier given by an inverted harmonic potential
(inverted parabola) with \bea U( q)=-{m\omega^2\over 2}
q^2.\label{inv}\eea The Hamiltonian $H_0$ in equation (\ref{ham})
with the potential (\ref{inv}) can be regarded as the Hamiltonian of
a harmonic oscillator with an imaginary frequency $i\omega$ and the
equations of motion (\ref{eqmo1})--(\ref{eqmo5}) for this potential
are formally obtained by performing the replacement $\omega\to
i\omega$ in the corresponding equations for the harmonic oscillator.
These equations of motions can be solved by using the same method as
in references \cite{rev,ss} for the harmonic oscillator. Contrary to
the situation of the harmonic oscillator, where we have two cases,
overdamped and underdamped, for the inverted parabola such a
distinction does not exist. The solutions for the expectation values
and variances of coordinate and momentum coincide formally with the
solutions corresponding to the overdamped case of the harmonic
oscillator. For the expectation value of the coordinate and momentum
we obtain with $\nu\equiv\sqrt{\omega^2+\mu^2}:$
\bea\sigma_q(t)=e^{-\lambda t}((\cosh\nu t+{\mu\over\nu}\sinh\nu
t)\sigma_q(0)+ {1\over m\nu}\sinh\nu t\sigma_p(0)), \label{sol6}\eea
\bea\sigma_p(t)=e^{-\lambda t}({m\omega^2\over\nu} \sinh\nu
t\sigma_q(0)+ (\cosh\nu t-{\mu\over\nu}\sinh\nu t)\sigma_p(0)).
\label{sol7}\eea In the following we also need the solution for the
variance $\sigma_{qq}$, which is given by: \bea
\sigma_{qq}(t)={e^{-2\lambda t}\over 2\nu^2}
\{\Delta_{qq}[(\mu^2+\nu^2)\cosh 2\nu t+2\mu\nu\sinh 2\nu
t+\omega^2] +{\Delta_{pp}\over m^2}(\cosh 2\nu t-1)\nonumber\\
+{2\over m}\Delta_{pq} (\mu\cosh 2\nu t+\nu\sinh 2\nu
t-\mu)\}+\sigma_{qq}(\infty),~~~~~~~~~~~~~~~~\label{qvar}\eea where
$\Delta_{qq}=\sigma_{qq}(0)-\sigma_{qq}(\infty),$
$\Delta_{pp}=\sigma_{pp}(0)-\sigma_{pp}(\infty),$
$\Delta_{pq}=\sigma_{pq}(0)-\sigma_{pq}(\infty)$ and
\bea\sigma_{qq}(\infty)={1\over
2m^2\lambda(\lambda^2-\omega^2-\mu^2)}
(m^2(2\lambda(\lambda+\mu)-\omega^2)D_{qq}
+D_{pp}+2m(\lambda+\mu)D_{pq}),\label{varinfqq}\eea
\bea\sigma_{pp}(\infty)={1\over
2\lambda(\lambda^2-\omega^2-\mu^2)}((m\omega)^2
\omega^2D_{qq}+(2\lambda(\lambda-\mu)-\omega^2)D_{pp}+2m\omega^2(\lambda-
\mu)D_{pq}),\label{varinfpp}\eea \bea\sigma_{pq}(\infty)={1\over
2m\lambda(\lambda^2-\omega^2-\mu^2)}((\lambda+
\mu)(m\omega)^2D_{qq}+(\lambda-\mu)D_{pp}+2m(\lambda^2-\mu^2)D_{pq}).
\label{varinfPQ}\eea Note that in the case $\lambda>\nu,$
$\sigma_q(t\to\infty)=\sigma_p(t\to\infty)=0$ and
$\sigma_{qq}(t\to\infty)=\sigma_{qq}(\infty).$ In the case
$\lambda<\nu$, $\sigma_q(t\to\infty),
\sigma_p(t\to\infty)\to\pm\infty$ and
$\sigma_{qq}(t\to\infty)=\infty.$

\section{Tunneling through an inverted parabola}

In order to calculate the tunneling probability through the inverted
harmonic potential (\ref{inv}), we assume that initially the wave
function of the system is a Gaussian wave packet centered at the
left of the peak of the potential at $q=0,$ $\sigma_q(0)<0$, with a
momentum $\sigma_p(0)>0$ towards the potential barrier peak: \bea
\psi(q)={1\over (2\pi\sigma_{qq}(0))^{1/4}}\exp[-{1\over
4\sigma_{qq}(0)} (q-\sigma_q(0))^2+{i\over\hbar}\sigma_p(0)q].\eea
Then the corresponding initial probability density is given by: \bea
\rho(q,t=0)={1\over (2\pi\sigma_{qq}(0))^{1/2}} \exp[-{1\over
2\sigma_{qq}(0)} (q-\sigma_q(0))^2].\eea

Like in \cite{a}--\cite{vlas}, we can transform the master equation
(\ref{mast}) for the density operator of a particle moving in the
potential (\ref{inv}) of an inverted parabola into the following
Fokker-Planck equation satisfied by the Wigner distribution function
$W(q,p,t):$ \bea   {\partial W\over\partial t}= -{p\over m}{\partial
W\over\partial q} -m\omega^2 q{\partial W\over\partial p}
+(\lambda-\mu){\partial\over\partial q}(qW)
+(\lambda+\mu){\partial\over\partial p}(pW) \nonumber \\
+D_{qq}{\partial^2 W\over\partial q^2} +D_{pp}{\partial^2
W\over\partial p^2} +2D_{pq}{\partial^2 W\over\partial p\partial
q}.~~~~~~~~~~~~~~~~~~~ \label{wigeq}\eea For an initial Gaussian
Wigner function, the solution of Eq. (\ref{wigeq}) is \bea
W(q,p,t)={1\over 2\pi\sqrt{\sigma(t)}}
~~~~~~~~~~~~~~~~~~~~~~~~~~~~~~~~~ \nonumber \\
\times\exp\{-{1\over 2\sigma(t)}[\sigma_{pp}(t)(q-\sigma_q(t))^2+
\sigma_{qq}(t)(p-\sigma_p(t))^2-2\sigma_{pq}(t)(q-\sigma_q(t))
(p-\sigma_p(t))]\},\label{wig} \eea which represents the most
general mixed squeezed states of Gaussian form. Here $\sigma_q(t),$
$\sigma_p(t)$ and $\sigma_{qq}(t),$ $\sigma_{pp}(t),$
$\sigma_{pq}(t)$ are the expectation values and, respectively, the
variances corresponding to the inverted parabola as given partly in
Eqs. (\ref{sol6})--(\ref{qvar}) and \bea
\sigma(t)=\sigma_{qq}(t)\sigma_{pp}(t)-\sigma_{pq}(t)^2.\label{det}\eea
Since the dynamics is quadratic, then according to known general
results, the initial Wigner function remains Gaussian. The density
matrix can be obtained by the inverse Fourier transform of the
Wigner function: \bea <q|\rho|q'>=\int dp \exp({i\over
\hbar}p(q-q'))W({q+q'\over 2},p,t).\label{fourinv}\eea Using Eq.
(\ref{wig}), we get for the density matrix the following time
evolution: \bea <q|\rho|q'>=({1\over 2\pi\sigma_{qq}(t)})^{1\over 2}
\exp[-{1\over
2\sigma_{qq}(t)}({q+q'\over 2}-\sigma_q(t))^2 ~~~~~~~~~\nonumber \\
-{1\over 2\hbar^2}
(\sigma_{pp}(t)-{\sigma_{pq}^2(t)\over\sigma_{qq}(t)})(q-q')^2
+{i\sigma_{pq}(t)\over\hbar\sigma_{qq}(t)}({q+q'\over
2}-\sigma_q(t))(q-q')+ {i\over
\hbar}\sigma_p(t)(q-q')].\label{ccd}\eea The initial Gaussian
density matrix also remains Gaussian, centered around the classical
path, i. e. $\sigma_q(t)$ and $\sigma_p(t)$ give the average
time-dependent location of the system along its trajectory in phase
space. The wave function starts as a Glauber wave packet at $t=0$ on
the left-hand side of the barrier and evolves as a mixed squeezed
state at a later time. By putting $q'=q$ in Eq. (\ref{ccd}), we
obtain the following probability density of finding the particle in
the position $q$ at the moment $t$: \bea \rho(q,t)={1\over
(2\pi\sigma_{qq}(t))^{1/2}} \exp[-{1\over 2\sigma_{qq}(t)}
(q-\sigma_q(t))^2].\label{dens}\eea This is a Gaussian distribution
centered at $\sigma_q(t),$ which describes the classical trajectory
of a particle initially at $\sigma_q(0)$, with initial momentum
$\sigma_p(0)$ and variance $\sigma_{qq}(t).$

Using Eq. (\ref{dens}), the probability for the particle
to pass to the right of position $q$ at time $t$ is given
by \bea P(q,t)=\int_q^\infty \rho(q',t)dq'=\int_q^\infty
{1\over\sqrt{2\pi\sigma_{qq}(t)}}\exp(-{(q'-\sigma_q(t))^2\over
2\sigma_{qq}(t)})dq'.\eea We define the tunneling
probability $P(t)$ as the probability for the particle to
be at the right of the peak at $q=0$ (beyond the barrier
top): $P(t)=P(q=0,t).$ We obtain \bea P(t)={1\over\sqrt{
\pi}} \int_{-\sigma_q(t)\over
\sqrt{2\sigma_{qq}(t)}}^\infty e^{-u^2}du ={1\over
2}(1-{\rm erf}(-{\sigma_q(t)\over\sqrt
{2\sigma_{qq}(t)}})), \label{probt}\eea where $ {\rm
erf}(x)$ is the error function with ${\rm erf}(x)={-\rm
erf}(-x)$ and erf($\infty$)=1.

From Eq. (\ref{probt}) we see that the probability $P(t)$
depends only upon the classical motion of the average
value of coordinate (wave packet center) and the
spreading of the wave packet in the direction of the
barrier. The final tunneling probability (barrier
penetrability) is given by taking the limit $t\to\infty$
in $P(t)$. In the present calculations we ignore the fact
that a part of the wave packet has already tunneled
through the barrier at $t=0.$ In general, this
probability has a negligible value, but, in principle, in
order to find the net penetration probability, it should
be subtracted from the tunneling probability at time $t.$

\section{Evaluation of the penetration probability and \\
analysis in dimensionless variables}

We will show that since $\sigma_q(t)$ and
$\sqrt{\sigma_{qq}(t)}$ are both proportional to the same
exponential factor as time approaches infinity, their
ratio in Eq. (\ref{probt}) approaches a finite limit,
which determines the final tunneling probability. Indeed,
as $t\to\infty,$ we see from Eqs. (\ref{sol6}) and
(\ref{qvar}) that $\sigma_q(t)$ and $\sigma_{qq}(t)$
behave like \bea\sigma_q(t)\to {e^{-(\lambda-\nu)t}\over
2m\nu}\delta\eea and \bea\sigma_{qq}(t)\to
({e^{-(\lambda-\nu)t}\over
2m\nu})^2\Delta+\sigma_{qq}(\infty),\eea where we have
denoted \bea \delta\equiv m
(\mu+\nu)\sigma_q(0)+\sigma_p(0)\eea and \bea
\Delta\equiv
m^2(\mu+\nu)^2\Delta_{qq}+\Delta_{pp}+2m(\mu+\nu)\Delta_{pq}.
\label{delta}\eea Then we obtain the following finite
limit as $t\to\infty:$ \bea
{\sigma_q(t)\over\sqrt{\sigma_{qq}}}\to {\delta \over
\sqrt{\Delta}}\eea if $\lambda<\nu$ (and the limit $0$ if
$\lambda>\nu$) and, therefore, the expression
(\ref{probt}) leads to the final penetration probability
$(P=P(t\to\infty))$ \bea P={1\over 2}(1-{\rm
erf}(-{\delta\over \sqrt{2\Delta}}))\eea if $\lambda<\nu$
and $P=1/2$ if $\lambda>\nu.$ In the case $\lambda>\nu,$
$\sigma_q(t)\to 0$ if $t\to\infty$, that is, the system
is located around the barrier, $\sigma_{qq}(t)$ tends to
a finite value $\sigma_{qq}(\infty)$ for any initial
kinetic energy and in this case $P(t)=1/2.$ Let us
consider the other case $\lambda<\nu$. For $\delta=0$ the
trajectory tends to the top of the potential barrier and
if $\delta$ is different from 0, then $\sigma_q(t)$ tends
to $\infty$ or $-\infty.$ The trajectory which starts on
the left-hand side of the barrier ($\sigma_q(0)<0$) with
a positive initial momentum $\sigma_p(0)>0$ will stay on
the same side for $\delta<0$ (and then
$\sigma_q(\infty)\to -\infty$) and will cross the barrier
for $\delta>0,$ i. e. if the initial kinetic energy
allows to overcome the barrier (and then
$\sigma_q(\infty)\to\infty$). For a general $\mu$ and any
$\lambda<\nu,$ the particle crosses the barrier when
$\sigma_p(0)>-m(\mu+\nu)\sigma_q(0)$ and will stay on the
same side when $\sigma_p(0)<-m(\mu+\nu)\sigma_q(0).$ The
barrier penetrability is larger than 1/2 if the particle
classically can overcome the top of the barrier, it is
smaller than 1/2 if the particle cannot cross the barrier
and it tends to 1/2 if the position uncertainty
$\sigma_{qq}$ is very large.

For $\mu=0,$ the values of $\Delta$ and $\delta$ become
($\nu=\omega$): \bea
\delta_0=m\omega\sigma_q(0)+\sigma_p(0)\eea and \bea
\Delta_0=m^2\omega^2\sigma_{qq}(0)+\sigma_{pp}(0)+2m\omega\sigma_{pq}(0)-
{m^2\omega^2D_{qq}+D_{pp}+2m\omega D_{pq}\over
\lambda-\omega}.\eea The particle crosses the barrier if
$\sigma_ p(0)>-m\omega\sigma_q(0)$ and $\lambda<\omega.$
In this case, if $\lambda$ increases, then the ratio
$\delta_0/(2\Delta_0)^{1/2}$ and the penetration
probability $P$ decreases. This means that if dissipation
increases, then the probability $P$ decreases up to a
value of 1/2. If $\sigma_ p(0)<-m\omega\sigma_q(0),$ the
particle can not cross the barrier. In this case the
penetration probability $P$ increases with the
dissipation $\lambda$ up to 1/2. At $\lambda\ge \omega$
the wave packet sticks in the barrier region.

We now introduce the dimensionless variables: $z$ the scaled initial
position, $v$ the scaled initial momentum, $\epsilon$ the scaled
dissipation coefficient and $r$ the scaled inverse wave packet size,
defined as follows: \bea z={\sigma_q(0)\over \sqrt{\sigma_{qq}(0)}},
~~v={\sigma_p(0)\over m\omega\sigma_q(0)},
~~\epsilon={\lambda\over\omega}, ~~r={\sqrt{\displaystyle{\hbar\over
2m\omega}}\over\sqrt{\sigma_{qq}(0)}}.\label{not}\eea With these
notations and considering a thermal bath modeled by the coefficients
of the form (\ref{coegib}), the penetration probability takes the
following form for the case $\mu=0$: \bea P_{\mu=0}={1\over
2}(1-{\rm erf}(-{\delta_0\over\sqrt{2\Delta_0}}))= {1\over 2}(1-{\rm
erf}(-{z(1+v)\over\sqrt{2}
\displaystyle{\sqrt{1+r^4-{2\epsilon\over\epsilon-1}r^2\coth{\hbar\omega
\over 2kT}}}})).\eea We took into account that \bea
\sigma_{qq}(0)\sigma_{pp}(0)=\hbar^2/4, ~~~~(\sigma_{pq}(0)=0).\eea
If $\mu=0,$ then we have $0<\epsilon<1.$

For $\mu\neq 0,$ in the case of a thermal bath, the
expression (\ref{delta}) takes the form
$(\sigma_{pq}(0)=0)$ \bea
\Delta=m^2(\mu+\nu)^2\sigma_{qq}(0)+\sigma_{pp}(0)\nonumber\\
-{\hbar
m\over\omega}\{\mu(\mu+\nu)+{\omega^2[\lambda^2+\lambda(\mu+\nu)+\mu\nu]\over
\lambda^2-\omega^2-\mu^2}\}\coth{\hbar\omega\over 2kT}.\eea With the
notations (\ref{not}) and introducing also the notation
\bea\gamma={\mu\over\omega},\eea the penetration probability takes
the following form: \bea P_{\mu\neq 0} ={1\over2}(1-{\rm
erf}(-{\delta\over\sqrt{2\Delta}})),\eea with \bea
{\delta\over\sqrt{\Delta}}={z(\gamma+\sqrt{1+\gamma^2}+v)\over
[(\gamma+\sqrt{1+\gamma^2})^2+r^4-2(\gamma+\sqrt{1+\gamma^2})
\displaystyle{{{(\epsilon^2-\gamma^2)\sqrt{1+\gamma^2}+\epsilon
\over\epsilon^2-\gamma^2-1}}r^2\coth{\hbar\omega\over
2kT}}]^{1/2}}.\eea If $\mu\neq 0$, the inequalities $\lambda>\mu$
(see Eq. (\ref{coegib}) and $\lambda<\nu$ lead to the following
restrictions on the dimensionless variables: \bea
\gamma<\epsilon<\sqrt{1+\gamma^2}.\eea The initial energy of the
particle associated with the Gaussian wave packet (21) is
\begin{equation} E=<H>|_{t=0}={1\over 2m}\sigma_{pp}(0)-{m\omega^2\over 2}
\sigma_{qq}(0)+{1\over 2m}\sigma^2_{p}(0)-{m\omega^2\over
2}\sigma^2_{q}(0)+ \mu\sigma_p(0)\sigma_q(0)\end{equation} and in
terms of the dimensionless variables (39) it looks
\begin{equation} E={\hbar\omega\over 4r^2}[r^4-1+z^2(v^2-1)]+{\hbar\mu\over
2}{z^2v\over r^2}.\end{equation} If $E<0$ it is a sub-barrier
initial energy and if $E>0$ it is an energy above the barrier. In
terms of the same dimensionless variables, the condition that a
classical particle does not have enough initial kinetic energy to
pass the potential barrier can be written:
\begin{equation} v>-1,~~~~~ {\rm if}~ \mu=0\end{equation}
and
\begin{equation} v>-(\gamma+\sqrt{1+\gamma^2}),~~~~~ {\rm if}~ \mu\not= 0.
\end{equation}
With these two conditions and by taking $0<r\leq 1$ (which assures
that the initial fluctuation energy is negative), the total initial
energy $E$ (47) is always negative. This corresponds to the case of
the sub-barrier energy, relevant to the quantum tunneling problem.
The examples provided in the following figures reflect just this
situation.

For $\mu=0,$ Figs. 1 and 2 show the dependence of the tunneling
probability on the scaled dissipation $\epsilon$ and the temperature
$T$ of the thermal bath, via $\coth({\hbar\omega/2kT}),$ for fixed
values of the scaled initial position $z$, momentum $v$ and wave
packet size $r.$

\begin{figure}
\label{Fig1}
\centerline{\epsfig{file=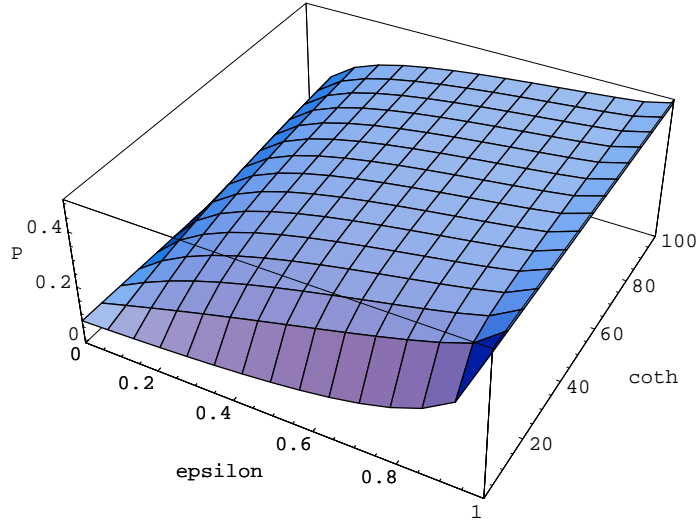,width=0.6\textwidth}}
\caption{Dependence of tunneling probability P on the
scaled dissipation $\epsilon={\lambda/\omega}$ and the
temperature $T$ of the thermal bath, via
$\coth({\hbar\omega/2kT}),$ for $\mu=0$ and for fixed
values of the scaled initial position $z=-3,$ scaled
initial momentum $v=-0.5$ and scaled inverse wave packet
size $r=0.5.$ }
\end{figure}

\begin{figure}
\label{Fig2}
\centerline{\epsfig{file=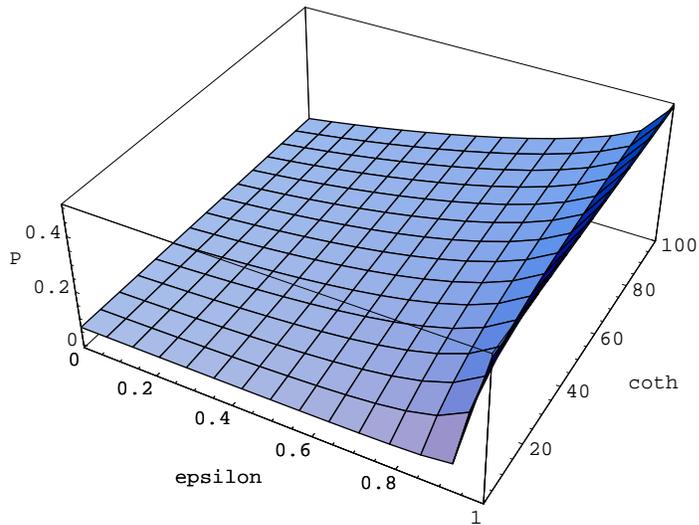,width=0.6\textwidth}}
\caption{Same as in Fig. 1, but with $z=-3,$ $v=-0.5$ and
$r=0.1.$ }
\end{figure}

In the next four figures, we consider $\mu\neq 0.$ Figs. 3 and 4
show the dependence of the penetration probability on the scaled
dissipation and on the parameter $\gamma$ for a fixed scaled initial
position $z$, momentum $v$ and wave packet size $r$ at the
temperature $T=0.$

\begin{figure}
\label{Fig3}
\centerline{\epsfig{file=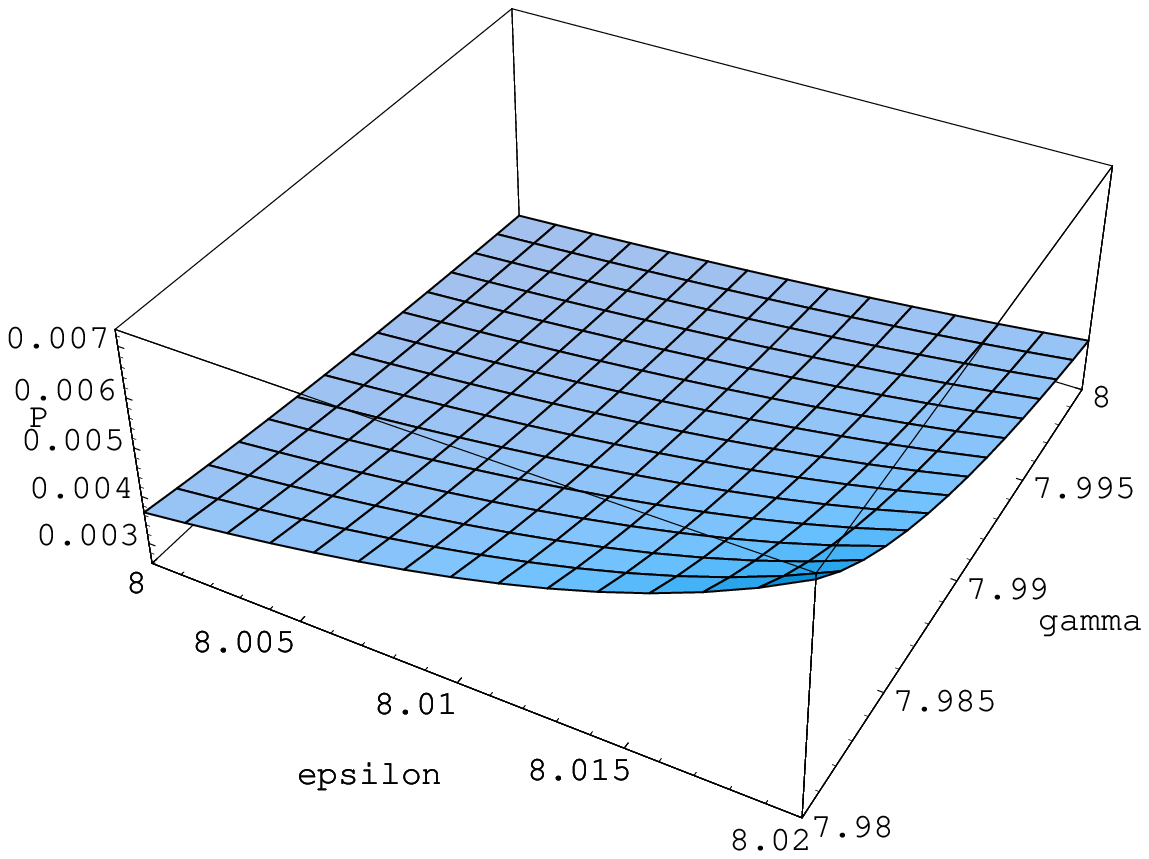,width=0.6\textwidth}}
\caption{Dependence of tunneling probability P on the
scaled dissipation $\epsilon$ and parameter
$\gamma=\mu/\omega,$ for the temperature $T=0$ of the
thermal bath and for fixed values of the scaled initial
position $z=-3,$ scaled initial momentum $v=-0.5$ and
scaled inverse wave packet size $r=0.3.$}
\end{figure}

\begin{figure}
\label{Fig4}
\centerline{\epsfig{file=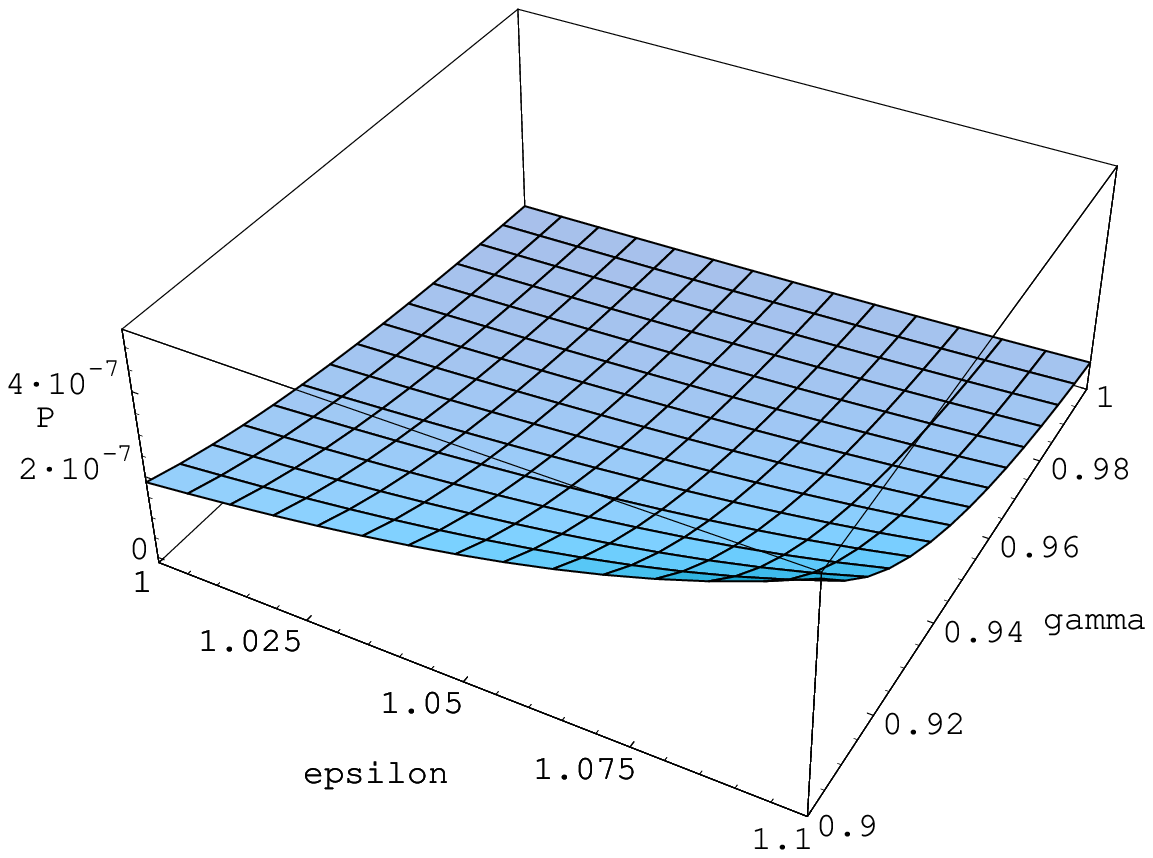,width=0.6\textwidth}}
\caption{Same as in Fig. 3, but with $z=-9,$ $v=-0.9$ and
$r=0.3.$ }
\end{figure}

In Figs. 5 and 6 we give the dependence of the penetration
probability on the scaled dissipation and temperature, at fixed
values of $z, v, r$ and $\gamma.$ The presented dependence of the
penetration probability on these variables can be summarized in the
following conclusions:

\begin{figure}
\label{Fig5}
\centerline{\epsfig{file=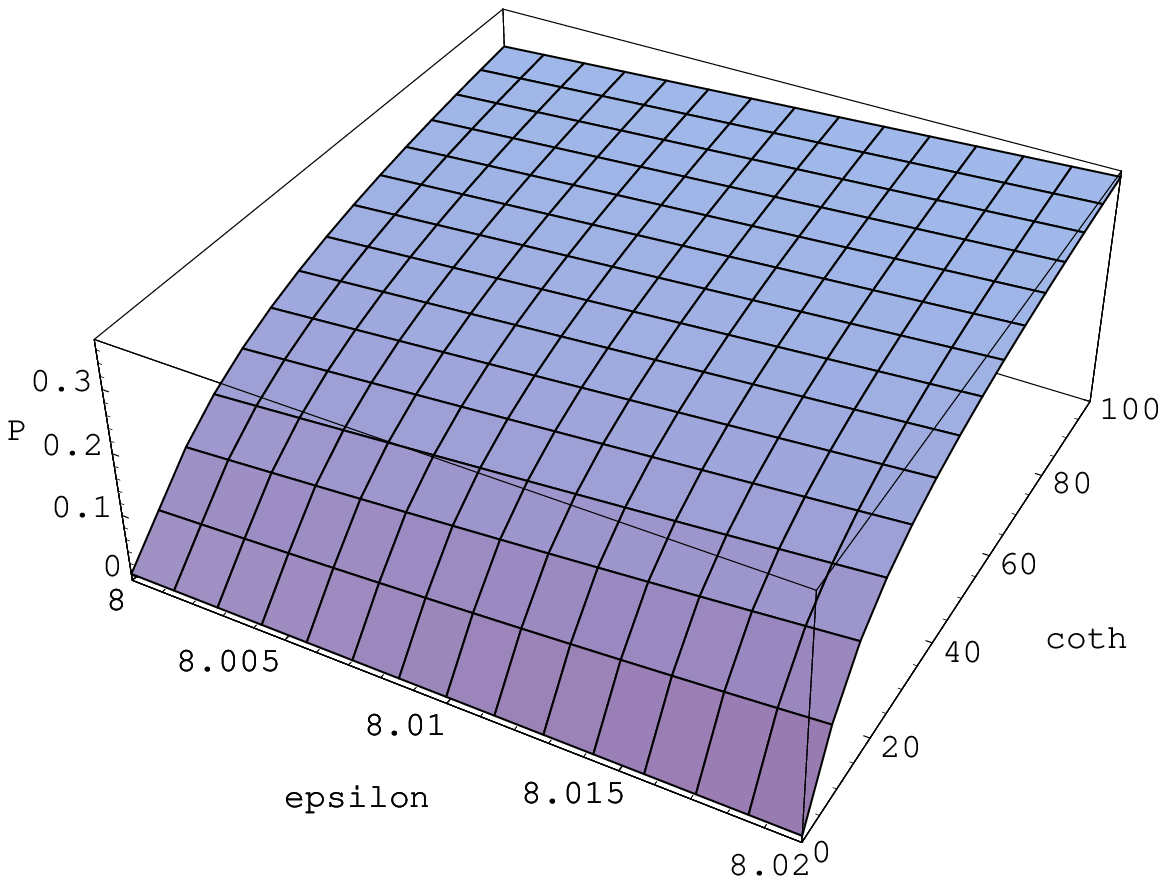,width=0.6\textwidth}}
\caption{Dependence of tunneling probability P on the
scaled dissipation $\epsilon$ and the temperature $T$ of
the thermal bath, via $\coth({\hbar\omega/2kT}),$ for
$\gamma=7.99$ and for fixed values of the scaled initial
position $z=-3,$ scaled initial momentum $v=-0.5$ and
scaled inverse wave packet size $r=0.5.$}
\end{figure}

\begin{figure}
\label{Fig6}
\centerline{\epsfig{file=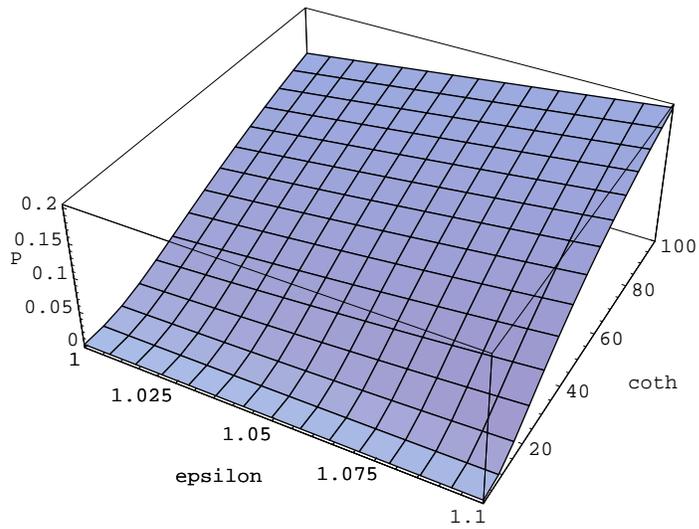,width=0.6\textwidth}}
\caption{Same as in Fig. 5, but with $\gamma=0.97,$
$z=-9,$ $v=-0.9$ and $r=0.5.$ }\end{figure}

1) When the scaled initial momentum $|v|$ is increasing, then $P$ is
increasing up to 1/2 if the particle does not have enough kinetic
energy to pass the potential barrier. The same conclusion is valid
for the variable $r,$ i. e. if the initial width of the Gaussian
packet is decreasing, then the penetration probability is
increasing.

2) If the scaled initial position $|z|$ is increasing, then $P$ is
decreasing from 1/2 to 0 if the particle does not have enough
kinetic energy to pass the potential barrier.

3) The penetration probability is increasing from 0 to 1/2 with
dissipation and with $\coth(\hbar\omega/2kT)$ and, therefore, with
the temperature. For the case $\mu\neq 0,$ the probability $P$ is
decreasing with $\mu.$

In conclusion, the dependence of the tunneling probability on
dissipation is not simple. When the particle does not have enough
kinetic energy to pass the parabolic barrier, which is the
relevant case to the quantum tunneling problem, the dissipation
enhances tunneling.

\section{Summary}

In the framework of the Lindblad theory for open quantum
systems, we have formulated the motion and the spreading
of Gaussian wave packets in an inverted oscillator
potential. We have obtained analytic solutions of
evolution in time of the wave packets and of the barrier
penetrability. Since the wave packets spread in time
according to the same law of evolution as their center
moves, the value of barrier penetrability is in general
different from 1/2. The inverted oscillator potential has
an important physical relevance, since it can constitute
a guide how to treat more physically realistic
potentials, like third order and double-well potentials
\cite{anto,anto1} or joined inverted parabola and
harmonic oscillator potentials \cite{misi}, in order to
be applied in nuclear fission and in molecular or solid
state physics.

{\bf Acknowledgments}

One of us (A. I.) is pleased to express his sincere gratitude for
the hospitality at the Institut f\"ur Theoretische Physik in
Giessen. A. I. also gratefully acknowledges financial support by
the DAAD (Germany).

\end{document}